


\font\twelvebf=cmbx10 scaled\magstep 1
\font\twelverm=cmr10 scaled\magstep 1
\font\twelveit=cmti10 scaled\magstep 1

\font\tenbf=cmbx10
\font\tenrm=cmr10
\font\tenit=cmti10

\parindent=1.5pc
\hsize=6.0truein
\vsize=8.5truein

\def\np#1#2#3{{\twelveit Nucl. Phys.} {\twelvebf B#1} (#2) #3}
\def\pl#1#2#3{{\twelveit Phys. Lett.} {\twelvebf #1B} (#2) #3}
\def\prl#1#2#3{{\twelveit Phys. Rev. Lett.} {\twelvebf #1} (#2) #3}
\def\physrev#1#2#3{{\twelveit Phys. Rev.} {\twelvebf D#1} (#2) #3}

\def\prep#1#2#3{{\twelveit Phys. Rep.} {\twelvebf #1} (#2) #3}

\def\cmp#1#2#3{{\twelveit Comm. Math. Phys.} {\twelvebf #1} (#2) #3}
\def\mpl#1#2#3{{\twelveit Mod. Phys. Lett.} {\twelvebf #1} (#2) #3}

\font\zfont = cmss10 
\font\litfont = cmr6

\def\bigone{\hbox{1\kern -.23em {\rm l}}}
\def\ZZ{\hbox{\zfont Z\kern-.4emZ}}
\def\half{{\litfont {1 \over 2}}}

\def\CM{{\cal M}}
\def\CF{{\cal F}}
\def\CA{{\cal A}}

\def\Im{{\rm Im ~}}

\def\Tr{{\rm Tr ~}}
\def\lfm#1{\medskip\noindent\item{#1}}
\overfullrule=0pt

\twelverm\baselineskip=14pt
\hfill{hep-th/9408013, RU-94-64, IASSNS-HEP-94/57}\bigskip
\baselineskip=22pt
\centerline{\bf THE POWER OF HOLOMORPHY -- }
\baselineskip=16pt
\centerline{\tenbf EXACT RESULTS IN 4D SUSY FIELD THEORIES}
\vglue 0.8cm
\centerline{\tenrm NATHAN SEIBERG}
\baselineskip=13pt
\centerline{\tenit Department of Physics and Astronomy, Rutgers
University}
\baselineskip=12pt
\centerline{\tenit Piscataway, NJ 08855-0849, USA}
\baselineskip=13pt
\centerline{\tenit and}
\baselineskip=13pt
\centerline{\tenit Institute for Advanced Study}
\baselineskip=12pt
\centerline{\tenit Princeton, NJ 08540, USA}
\vglue 0.8cm
\centerline{\tenrm ABSTRACT}
\vglue 0.3cm
{\rightskip=3pc \leftskip=3pc \tenrm\baselineskip=12pt\noindent
Holomorphy of the superpotential and of the coefficient of the gauge
kinetic terms in supersymmetric theories lead to powerful results.  They
are the underlying conceptual reason for the important
non-renormalization theorems.  They also enable us to study the exact
non-perturbative dynamics of these theories.  We find explicit
realizations of known phenomena as well as new ones in four dimensional
strongly coupled field theories.  These shed new light on confinement
and chiral symmetry breaking.
This note is based on a talk delivered at the PASCOS (94) meeting at
Syracuse University.
\vglue 0.6cm}

\vfil
\twelverm
\baselineskip=14pt
\leftline{\twelvebf 1. Introduction}
\vglue 0.4cm

Perhaps the most remarkable property of supersymmetric field theories is
the non-renormalization theorem.  In this talk we will give a conceptual
understanding of the theorem, and extend it beyond perturbation theory.
We will identify the holomorphy of the superpotential and of the
coefficient of the gauge kinetic term as the crucial elements in this
theorem$^1$.  The non-perturbative understanding will enable us to
derive exact results in these strongly coupled four dimensional
supersymmetric gauge theories$^{2,3,4}$ (see also, a related discussion
in reference 5).

There are three main applications to these supersymmetric theories:

\lfm{1.} The gauge hierarchy problem:  This was the original motivation
for studying the dynamics of supersymmetric field theories.  First, the
perturbative non-renor-malization protects a tree level hierarchy.
Non-perturbative violation of the perturbative theorem allows dynamical
supersymmetry breaking at low energies$^6$.  The new non-renormalization
theorem enables one to analyze the theory exactly.

\lfm{2.} Topological Field Theories$^7$.  We will have little to
say about them here except to note that we expect our new dynamical
insights to be helpful in understanding them.

\lfm{3.} The new application, which we will suggest here, is that they
can be used as ``toy models'' for the dynamics of strongly coupled four
dimensional field theories.  Our ability to find exact results allows us
to study long standing issues like chiral symmetry breaking and
confinement.

\bigskip

For lack of time, we will be rather brief and will mention only the main
points here.  The talk will be like a quick tour through the different
models and phenomena that were found.  The interested reader is referred
to the original papers for details.

\vglue 0.6cm
\leftline{\twelvebf 2.  Non-renormalization Theorem}
\vglue 0.4cm
\leftline{\twelveit 2.1. General Rules}
\vglue 0.4cm

The standard proof$^8$ of the non-renormalization theorem is based on
the study of Feynman diagrams in superspace.  Its main drawbacks are its
complexity and the limitation to perturbation theory.  The new proof
which we will present here generalizes observations of references 9, 10
and 11.  It is similar in spirit to the non-renormalization in sigma
model perturbation theory$^{12}$ and in semiclassical perturbation
theory$^{13}$ in string theory.

We will think of all the coupling constants in the tree level
superpotential $W_{\rm tree}$, $\lambda_I$, and the gauge coupling,
${4\pi\over g^2}\sim \log \Lambda$, as background fields.  Then, the
renormalized, effective superpotential, $W_{\rm eff}(\phi_i, \lambda_I,
\Lambda)$ is constrained by:

\lfm{1.} Symmetries: By assigning transformations laws both to the
fields and to the coupling constants the theory has a larger symmetry.
The effective Lagrangian should be invariant under it.  Such
restrictions are common in physics and are usually referred to
as selection rules.

\lfm{2.} Holomorphy: $W_{\rm eff}$ is independent of $\lambda_I^\dagger$.
This is the key property.  Just as the superpotential is holomorphic in
the fields, it is also holomorphic in the coupling constants (the
background fields).  This is unlike the effective Lagrangian in
non-supersymmetric theories which is not subject to any holomorphy
restrictions.

\lfm{3.} Various limits: $W_{\rm eff}$ can be analyzed approximately at
weak coupling.  The singularities have physical meaning and can be
controlled.

\medskip

Often this enables us to determine $W_{\rm eff}$ completely.  The point
is that a holomorphic function (more precisely, a section) is determined
by its asymptotic behavior and singularities.

A crucial subtlety has to be mentioned at this point.  As explained by
Shifman and Vainshtein$^{10}$ there are two different objects which are
usually called ``the effective action.''  The 1PI effective action and
the Wilsonian one.  When there are no interacting massless particles,
these two effective actions are identical.  However, when interacting
massless particles are present, the 1PI effective action suffers from IR
ambiguities and might suffer from holomorphic anomalies.  These are
absent in the Wilsonian effective action.

There are two obvious extensions to our discussion.  First, it easily
extends to the coefficient $f(\Phi)$ in $ \int d^2 \theta f(\Phi)
W_\alpha^2$ which is also holomorphic.  Second, in $N=2$ supersymmetry
the Kahler potential is related by supersymmetry to such an $f(\Phi)$
and therefore it is also constrained by our considerations.

\vglue 0.3cm
\leftline{\twelveit 2.2. Example: Wess-Zumino Model}
\vglue 0.4cm

In order to demonstrate our rules we study here the simplest Wess-Zumino
model and rederive the known non-renormalization theorem.  Consider the
theory based on the tree level superpotential
$$ W_{\rm tree} = m \phi^2 + \lambda \phi^3 . $$
We will make use of two $U(1)$ symmetries.  The charges of the field
$\phi$ and the coupling constants $m$ and $\lambda$ are

$$\matrix{\quad& U(1)& \times & U(1)_R \cr
\phi & 1 && 1 \cr
m &  - 2 && 0 \cr
\lambda & -3 && -1 \cr} $$
where $U(1)_R$ is an R symmetry.  Note that non-zero values for
$\lambda$ and $m$ explicitly break both of them.  However, the symmetry
still leads to selection rules.

The symmetries and holomorphy of the effective superpotential restrict
it to be of the form
$$ W_{\rm eff}= m \phi^2 f\left( {\lambda \phi \over m} \right). $$

We proceed, according to our general rules above, by studying the limit
of small $\lambda$.  In this region perturbation theory is valid and the
superpotential can be expanded as
$$ W_{\rm eff}= \sum_{n=0}^\infty a_n{1 \over m^{n-1}} \lambda^n
\phi^{n+2}.$$

The $n$'th term in this expansion can arise from a tree diagram with $n+2$
external legs, $n$ vertices and $n-1$ propagators.  For $n \ge 2$ it is
not 1PI and its contribution should not be included in the effective
action.  Higher order corrections in $\lambda$ with the same number of
external legs could arise from loop diagrams.  However, they must be
absent because they are not compatible with the form of $W_{\rm eff}$.

We conclude that the effective superpotential is
$$W_{\rm eff}= m \phi^2 + \lambda \phi^3 = W_{\rm tree};$$
i.e.\ the superpotential is not renormalized.

Thus, we rederive the standard perturbative non-renormalization theorem.
Furthermore, our result extends it beyond perturbation theory.  Strictly
speaking, the Wess-Zumino model probably does not exist as an
interacting quantum field theory in four dimensions and therefore this
non-perturbative result is of little interest.  However, such a model
does exist in two dimensions and there our non-perturbative result
applies.

If there are several fields, some of them are heavy and the other are
light, the heavy fields can be integrated out to yield a low energy
effective Lagrangian for the light fields.  Then, the contribution of
tree diagrams with intermediate heavy fields should be included in the
effective action.  As we saw, our simple rules allow such diagrams to
contribute.  Our result is thus compatible with the known tree level
renormalization of the superpotential.

\vglue 0.6cm
\leftline{\twelvebf 3. Applications To $N=1$ SUSY Gauge Theories}
\vglue 0.4cm
\leftline{\twelveit 3.1 Supersymmetric QCD -- Classical Moduli Space}
\vglue 0.4cm

We now turn to more complicated models which are asymptotically free and
probably exist non-perturbatively.  Consider a supersymmetric $SU(N_c)$
gauge theory with $N_f$ quark flavors in the fundamental representation.
The elementary fields are gluons, quarks and antiquarks:
$$\eqalign{
W_\alpha & = \lambda_\alpha +\theta_\beta \sigma^{\mu\nu \beta}_\alpha
F_{\mu\nu} + \dots \cr
Q_i &= q_i + \theta \psi_i + \dots \qquad\qquad i, \tilde i = 1,...,N_f
\cr
\tilde Q^{\tilde i} &= \tilde q^{\tilde i} + \theta \tilde \psi^{\tilde
i} + \dots \cr }$$

An important property of these theories is that in the absence of mass
terms they have a continuum of classical ground states -- flat
directions.  These are directions in field space along which the
classical potential vanishes.  Unlike the situation with spontaneous
symmetry breaking, these are physically inequivalent ground states.  In
string theory such a continuum of inequivalent ground states is generic.
Borrowing the terminology from string theory, we will refer to the space
of classical ground states as the ``classical moduli space.''  More
explicitly, up to gauge and global symmetry transformations they are
given by:
$$ q=\tilde q= \pmatrix { a_1& & & & \cr
&a_2& & & \cr
& & .
& & \cr
& & & a_{N_f} &\quad \cr} $$
for $N_f < N_c$ and by
$$q=
\pmatrix { a_1& & & \cr
&a_2& & \cr
& & .  & \cr
& & & a_{N_c}\cr
& & & \cr
& & & \cr} ; \quad
\tilde q= \pmatrix {
\tilde a_1& &       &         \cr
   &\tilde a_2&     &         \cr
   &   & .   &         \cr
   &   &     & \tilde a_{N_c}\cr
   &   &     &         \cr
   &   &     &         \cr} $$
$$ |a_i|^2 - |\tilde a_i|^2= {\rm independent ~ of} ~ i $$
for $N_f \ge N_c$.

Already at the classical level we can integrate out the massive fields
and consider an effective Lagrangian for the massless modes.  Their
expectation values label the particular ground state we expand around,
and hence they are coordinates on the classical moduli space.  We will
call them  moduli.  The classical moduli space is not smooth.  Its
singularities are at the points of enhanced gauge symmetry.  For
instance, when $a_i=\tilde a_i=0$ for every $i$ the gauge symmetry is
totally unbroken.  Therefore, the low energy effective theory of the
moduli is singular there.  This should not surprise us.  At these
singular points there are new massless particles -- gluons.  An
effective Lagrangian without them is singular.  If we include them in
the low energy description, the Lagrangian is smooth.

In the next subsections we will see how this picture changes in the
quantum theory.  At large field strength, far from the classical
singularities the gauge symmetry is broken at a high scale.  Therefore,
the quantum theory is weakly coupled and semiclassical techniques are
reliable.  We expect the quantum corrections to the classical
picture to be small there.  On the other hand, at small field strength
the quantum theory is strongly coupled and the quantum corrections can
be large and cause dramatic modifications to the classical behavior.

\vglue 0.4cm
\leftline{\twelveit 3.2. $N_f < N_c$ -- No Vacuum}
\vglue 0.4cm

The first question to ask is whether the classical vacuum degeneracy can
be removed quantum mechanically.  For that to happen a superpotential
will have to be generated.  There is a unique invariant
superpotential$^{14}$
$$W_{\rm eff} = (N_c-N_f) {\Lambda^{3N_c-N_f \over N_c-N_f} \over (\det
\tilde Q Q )^{1 \over N_c-N_f}} $$
where $\Lambda$ is the dynamically generated scale of the theory.
The prefactor $N_c-N_f$ is a choice of normalization of $\Lambda$ (a
choice of subtraction scheme).
Therefore, if the vacuum degeneracy is removed, this particular
superpotential must be generated.  For $N_f \ge N_c$ this superpotential
does not exist (either the exponent diverges or the determinant
vanishes) and therefore the vacuum degeneracy cannot be lifted.  We will
return to this case in the next subsections.

For $N_f < N_c$ this superpotential is generated dynamically$^{6}$.  For
$N_f=N_c-1 $ it is generated by instantons and for $N_f < N_c-1$ by
gluino condensation.  (For related work on this model see also
references 9 and 15.)  This dynamically generated superpotential leads
to a potential which slopes to zero at infinity.  Therefore, the quantum
theory does not have a ground state.  We started with an infinite set of
vacua in the classical theory and ended up in the quantum theory without
a vacuum!

\vglue 0.4cm
\leftline{\twelveit 3.3. $N_f = N_c$ -- Quantum Moduli Space}
\vglue 0.4cm

As we said above, for $N_f \ge N_c$ the vacuum degeneracy is not lifted.
Therefore, the quantum theory also has a continuous space of
inequivalent vacua.  Since this space can be different than the
classical one, we will refer to it as the ``quantum moduli space.''  The
most interesting questions about it are associated with the nature of
its singularities.  Classically, the singularities were associated with
massless gluons.  Are there singularities in the quantum moduli space?
what are the massless particles in those singularities?

In order to answer these questions we start with the first case of $N_f
= N_c$ and give a gauge invariant description of the moduli space.
Consider the mesons, baryon and anti-baryon composite fields
$$\eqalign{
M_i^{\tilde i} =& \tilde Q^{\tilde i} Q_i \cr
B=& Q^{N_c} \cr
\tilde B=& \tilde Q^{N_c} .} $$
Classically
$$ \det M - \tilde B B =0$$
which follows from Bose statistics of $Q$ and $\tilde Q$.  It is easy to
see that the classical moduli space is the space of $M$, $B$ and
$\tilde B$ subject to this relation.

We will not give the derivation here but will simply assert that the
quantum moduli space is parameterized by the same fields but the
constraint is modified$^2$ to
$$ \det M - \tilde B B =\Lambda^{2N_c} \not=0 .$$

It is important that this space is smooth.  There are no singularities
in the quantum moduli space.  All the singularities have been smoothed
out.  (For a simple low dimensional example of how such a singularity
can be smoothed out consider an hyperboloid in a cone.  Far from the tip
of the cone the two spaces are similar, but the singularity at the tip
of the cone is smoothed out on the hyperboloid.)

Since there are no singularities on the quantum moduli space, the only
massless particles are the moduli.  The gluons are ``Higgsed'' in the
semiclassical region of large fields and are confined for small fields.
Note that there is a smooth transition from a region where a Higgs
description is more appropriate to a region where a confining
description is more appropriate.  This is possible because of the
presence of matter fields in the fundamental representation of the gauge
group$^{16}$.

Different points on the quantum moduli space exhibit different patterns
of global symmetry breaking.  For example at $M_i^{\tilde i} =\Lambda^2
\delta_i^{\tilde i},$ $B=\tilde B=0$ the symmetry is broken as
$$SU(N_f)_L\times SU(N_f)_R \times U(1)_B \times U(1)_R \rightarrow
SU(N_f)_V \times U(1)_B \times U(1)_R .$$
At $M=0,$ $B=- \tilde B = \Lambda^{N_c}$ the breaking pattern is
$$SU(N_f)_L\times SU(N_f)_R \times U(1)_B \times U(1)_R \rightarrow
SU(N_f)_L\times SU(N_f)_R  \times U(1)_R .$$
In both cases some of the moduli are Goldstone bosons of the broken
symmetry.  It is straightforward to check that the massless fermions
saturate the 'tHooft anomaly conditions for the unbroken symmetries.

\vglue 0.4cm
\leftline{\twelveit 3.4. $N_f = N_c+1$ -- Confinement Without Chiral
Symmetry Breaking}
\vglue 0.4cm

We now add another flavor to the previous case.  The classical moduli
space is again described by the mesons $M$, baryons $B$ and anti-baryons
$\tilde B$ subject to the constraints
$$ \eqalign{&
\det M \left( {1 \over M} \right)_{\tilde i}^i - \tilde B_{\tilde i}
B^i =0 \cr
&M_i^{\tilde i} B^i= \tilde B_{\tilde i} M_i^{\tilde i} .\cr} $$

Unlike the previous case, here the quantum moduli space can be shown to
be the same as the classical one$^2$.  Therefore, it is singular
and we should interpret the singularities in the quantum theory.

It turns out$^2$ that in the quantum theory all the degrees of
freedom in $M$, $B$ and $\tilde B$ are physical and they couple through
the superpotential
$$W_{\rm eff}= {1 \over \Lambda^{2N_c -1 }}(\tilde B_{\tilde i}
M_i^{\tilde i} B^i - \det M ). $$
The classical constraints appears as the equations of motion ${\partial
W_{\rm eff} \over \partial M}= {\partial W_{\rm eff} \over \partial B}=
{\partial  W_{\rm eff} \over \partial \tilde B}=0$.

The singularities, however, are interpreted differently than in the
classical theory.  At the point $M=\tilde B = B=0$ the global chiral
symmetry $$SU(N_f)_L\times SU(N_f)_R \times U(1)_B \times U(1)_R$$ is
unbroken and all the components of $M$, $B$ and $\tilde B$ are massless.
Again, it is easy to check that the 'tHooft anomaly conditions are
saturated.

We conclude that the spectrum at the origin of field space consists of
massless composite mesons and baryons and the chiral symmetry of the
theory is unbroken there.  This is confinement without chiral symmetry
breaking.  Again, we see a smooth transition$^{16}$ from the
semiclassical region where a Higgs description is more appropriate
to a strongly coupled region where a confining description is more
appropriate.

\vglue 0.4cm
\leftline{\twelveit 3.4. $N_f > N_c+1$ -- No Confinement; Interacting
Superconformal Theories}
\vglue 0.4cm

In this subsection we continue to add flavors and discuss the situation
of $N_f > N_c+1$.  As for $N_f = N_c+1$, the classical moduli space is
unchanged quantum mechanically$^2$.  Therefore, there are singularities
in the quantum moduli space and new massless states should appear there.
However, it can be shown that it is impossible to interpret the fields
which create them as gauge invariant polynomials in the elementary
fields.  Therefore, we conjecture that at the singularities the
fundamental quarks and the gluons are massless.

According to standard quantum field theory, the theory of the massless
states must be scale invariant.  In the previous cases this theory was
free.  Now, on the other hand, the massless fields interact, so the low
energy theory must be a non-trivial scale invariant theory in which the
beta function vanishes.

Such interacting scale invariant theories in four dimensions are known
to exist for large $N_c$ and $N_f$, by balancing the first two terms in
the beta function.  The resulting fixed point is at weak coupling thus
justifying the expansion.  More explicitly, we can show that this fixed
point exists for $N_f, N_c \rightarrow \infty$ with $\epsilon= {3N_c-N_f
\over N_c}$ small.  We conjecture that such a fixed point persists for
small $N_f$ and $N_c$ for every $N_f > N_c+1$.

We conclude that at the singular points on the moduli space the theory
is at non-trivial fixed points of the renormalization group; i.e.\
these are interacting superconformal field theories.

In such a scale invariant theory the beta function vanishes but the
operators can have non-canonical dimensions.  We cannot compute all
these dimensions.  However, for chiral superfields the dimensions are
determined as follows.  The superconformal algebra in four dimensions
includes a $U(1)_R$ symmetry.  The dimensions of chiral fields, $D$, are
related to their $R$ charge by
$$D={3 \over 2} R.$$
Around the UV (semiclassical) fixed point there is only one anomaly free
R symmetry which commutes with all the flavor symmetries.  Therefore,
this must be the symmetry in the superconformal algebra at the
non-trivial fixed point.  We conclude that$^{2, 17}$
$$D(\tilde Q Q) = 3{N_f-N_c \over N_f }. $$

\vglue 0.4cm
\leftline{\twelveit 3.5. Non-trivial Superpotentials}
\vglue 0.4cm

In the previous examples the superpotential and/or the structure of the
moduli space was essentially determined by the symmetries, holomorphy and
the weak coupling asymptotics.  Correspondingly, in all these models
$W_{\rm eff}$ is a single power of the fields which is fixed by the
symmetries.  More complicated models, where more powerful techniques are
needed, were studied in reference 3.

Consider, for a example, a model based on the gauge group $SU(2)_1
\times SU(2)_2$ with matter fields transforming like
$$\matrix{
Q &  {\bf( 2, 2 )} \cr
L_+ &  {\bf (1 ,2 )} \cr
L_- &  {\bf (1 ,2 )} .\cr}$$

The classical moduli space is two complex dimensional (at the generic
point in the moduli space the gauge group is completely broken, hence
six out of the eight elementary chiral
matter fields acquire a mass and the other two
remain massless).  It can be parametrized by the gauge invariant order
parameters
$$\eqalign{ X=& Q^2 \cr
Y=& L_+L_- .} $$
The classical singularities are at the points where either $X=0$ where
$SU(2)_1$ is unbroken, or $Y=0$ where a diagonal subgroup of $SU(2)_1
\times SU(2)_2$ is unbroken.

Symmetries and holomorphy constrain$^3$ the effective superpotential to
be of the form
$$ W_{\rm eff} = {\Lambda_1^5 \over X} f\left(
{\Lambda_2^4 \over XY} \right) $$
($\Lambda_1$ and $\Lambda_2$ are the dynamically generated scales
of the two gauge groups)
but do not determine the function $f$.  At weak coupling (large $XY$)
the gauge symmetry is completely broken and the superpotential can
be generated by instantons.  Hence
$$ W_{\rm eff} = {\Lambda_1^5 \over X} \sum_{n=0}^\infty a_n
\left( { \Lambda_2^4 \over XY}\right)^n $$
where the $n$'th term in the sum can arise from instantons with
instanton numbers $(1,n)$ under the two gauge groups.

A more detailed analysis is needed in order to determine the coefficients
$a_n$.  The main point$^3$ is that the classical singularity at $XY=0$
moves in the quantum theory to $XY=\Lambda_2^4$.  Examining the behavior
of $W_{\rm eff}$ near the point where the argument of $f\left(
{\Lambda_2^4 \over XY}\right) $ equals one leads to
$$ W_{\rm eff} = {\Lambda_1^5 Y \over XY - \Lambda_2^4} =
\sum_{n=0}^\infty {\Lambda_1^5 \over X} \left( { \Lambda_2^4 \over XY}
\right)^n .$$

We see that the combination of symmetries, holomorphy, and asymptotic
behavior at weak coupling and near the singularities enables us to sum
up the instanton series and to find a non-trivial answer.

\vglue 0.6cm
\leftline{\twelvebf 4. Applications To $N=2$ SUSY Gauge Theory}
\vglue 0.4cm

In the previous sections we demonstrated the power of holomorphy
in determining the superpotential and the topology of the quantum
moduli space.  However, we were unable to determine the metric
on the moduli space.  The reason for that is that the metric on
the space $(ds)^2=g_{i \bar j}d\phi^i d \bar \phi^{\bar j}$
is given by the kinetic term of the moduli
$$g_{i \bar j}(\phi, \bar \phi) \partial_\mu \phi^i \partial^\mu
\bar \phi^{\bar j}$$
which is derived from the Kahler potential
$$g_{i \bar j}(\phi, \bar \phi )=\partial_i \bar \partial_{\bar j}
K(\phi, \bar \phi). $$
Since $K$ is not holomorphic, it is not constrained by the previous
considerations.  However, in $N=2$ theories the Kahler potential is
controlled by a holomorphic function and therefore it can be
determined$^4$.

To be more specific, consider an $SU(2)$ gauge theory coupled to a
single chiral matter field, $\Phi$, in the adjoint representation.  This
theory has $N=2$ supersymmetry.

The classical moduli space is one complex dimensional corresponding to
the flat direction of the potential
$$\langle \Phi \rangle = \half \pmatrix{
a & 0 \cr
0 & -a}  . $$
It can be labeled by the gauge invariant order parameter $u=\langle {\Tr
\Phi^2}\rangle=\half a^2 $.  This expectation value breaks the $SU(2)$
gauge symmetry to $U(1)$.  Therefore, the light fields along this
direction are a photon and its $N=2$ superpartners -- a Dirac fermion
and a complex scalar.  The expectation value of this scalar $a$ is a
local coordinate on the classical moduli space.  $u= \half a^2$ is a
good global coordinate.

The low energy effective action is determined by a single holomorphic
function, the prepotential, $\CF(\CA)$,
$${1 \over 4\pi} \Im \left[ \int d^4 \theta
{\partial \CF(A)\over\partial A}
\bar A + \int d^2\theta \half  {\partial^2 \CF(A)\over\partial A^2}
W_\alpha W^\alpha \right]$$
where $A$ is an $N=1$ chiral multiplet and $W_\alpha$ is the $N=1$
field strength of a photon.  They combine into a single $N=2$ vector
multiplet $\CA$ whose scalar component is $a$.  This is the flat space
limit of ``special geometry.''$^{18}$

The leading quantum corrections to the metric on the moduli space were
studied in reference 19
$$\CF= i {1\over 2\pi} \CA^2\ln {\CA^2 \over \Lambda^2} +
\sum_{k=1}^\infty \CF_k \left( {\Lambda \over \CA} \right)^{4k}
\CA^2 $$
where the $k$'th term in the sum represents the contribution of $k$
instantons.

In reference 4 the exact theory was analyzed.  A crucial point is the
fact that the weak coupling description above is not valid globally.
The coordinate $a$ and the function $\CF$ are not single valued on
the moduli space.  Instead, the pair
$$\left(\eqalign{& A_D= {\partial \CF(A)\over\partial A}\cr &A}
\right)$$
is a section of an $SL(2,Z)$ bundle over the moduli space.  More
physically, there is freedom to perform duality transformations
on the light fields.  They are generated by
$$\eqalign{T&=\pmatrix{1 & 1 \cr 0 & 1} \cr
S&=\pmatrix{0 & 1 \cr -1 & 0} .\cr}$$
The $N=1$ gauge multiplet $W_\alpha$ is invariant under $T$ and
undergoes a standard electric-magnetic duality transformation under $S$.
The pair $(A_D= {\partial \CF(A)\over\partial A},\, A)$ transforms as a
column vector which is multiplied by the matrices $T$ or $S$.  These two
generators generate the group $SL(2,Z)$.  As we move around a closed
path on the moduli space, the light fields are not single valued;
they can be transformed by an $SL(2,Z)$ transformation.

At large field strength the semiclassical expression above shows that
$a_D \approx i {1\over \pi} a\ln a^2$ where $a_D$ is the scalar
component of the superfield $A_D$.  In terms of the gauge invariant
coordinate $u \approx \half a^2$ our pair is
$$\left(\eqalign{&a_D \approx  i {1\over \pi} \sqrt{2u} \ln u  \cr
&a \approx\sqrt{2u} }\right) . $$
As we move around the moduli space $u \rightarrow e^{2\pi i}u$ the pair
transforms by
$$\CM_\infty= PT^{-2}= \pmatrix{-1 & 2 \cr 0 & -1} $$
($P=-1$ is the charge conjugation matrix).  We see that the pair is
not single valued even in the weak coupling region.

It turns out$^4$ that the quantum moduli space is the complex $u$ plane
with two singular points at $u=\langle \Tr \Phi^2\rangle =\pm 1$ (after
rescaling the value of $u$ by a factor proportional to $\Lambda^2$).
The pair $(a_D,a)$ transforms by
$$ \CM_1=ST^2S^{-1}=\pmatrix{1 & 0 \cr -2 & 1}$$
as we circle around the singularity at $u=1$ and by
$$ \CM_{-1}=(TS)T^2(TS)^{-1}=\pmatrix{-1 & 2 \cr -2 & 3} $$
as we circle around $u=-1$.  Using this information the pair
$(a_D(u),a(u))$ can be found$^4$
$$\eqalign{
&a_D(u)={\sqrt 2 \over \pi} \int_1^u {dx \sqrt{x-u} \over \sqrt
{x^2 -1} }\cr
&a(u)={\sqrt 2 \over \pi} \int_{-1}^1 {dx \sqrt{x-u} \over
\sqrt {x^2 -1} }. } $$
The Kahler potential which is $K=\Im a_D(u)\bar a (\bar u)$ is therefore
also known and so is the metric derived from it.

As in our previous examples the singularities on the moduli space are
associated with new massless particles.  In this case these are massless
magnetic monopoles$^4$.  The low energy effective theory around these
points is non-local in terms of the elementary photon.  However, the
dual of the photon, ``the magnetic photon,'' couples locally to the
massless magnetic monopoles.  Therefore, the low energy theory looks
like $N=2$ supersymmetric QED with massless charged fields (the
monopoles).

The knowledge of $(a_D(u),a(u))$ enables us to find the masses of the
stable BPS-saturated$^{20}$ states.  As explained by Witten and
Olive$^{21}$, their masses are related to a central extension in the
$N=2$ algebra.  Including the quantum corrections the masses are
expressed in terms  of the magnetic and electric charges of the
particles $(n_m,n_e)$ as$^4$
$$M=\sqrt 2 |Z| \qquad {\rm with} \qquad Z=n_ma_D(u)+n_ea(u) . $$

We can break $ N=2$ supersymmetry to $N=1$ by adding a mass term to the
chiral superfield $\Phi$, $W=m\Tr \Phi^2$.  Then, the following things
happen$^4$:

\lfm{1.} The classical moduli space becomes one point $\Phi=0$ where
all the elementary fields are massless.  The quantum moduli space
collapses to the two singular points with $u= \langle \Tr \Phi^2
\rangle =\pm 1$.  Since the value of $u$ is non-zero at these points,
the chiral $Z_2$ symmetry of the theory is spontaneously broken.

\lfm{2.} The massless monopoles at these points condense.  This
condensation gives a mass to the previously massless photon by the Higgs
mechanism and a mass gap is generated.  However, since this photon is
dual to the electric photon, what we see here is actually confinement of
electric charges.

\medskip

For $m \gg \Lambda$, the elementary chiral field $\Phi$ can
be integrated out at high energies.  The low energy theory is the $N=1$
supersymmetric Yang-Mills theory without matter fields.  This theory is
expected to confine and to have two ground states with a mass gap in
which its discrete $Z_2$ chiral symmetry is spontaneously broken$^{22}$.
These are precisely the phenomena we saw above.

\vglue 0.6cm
\leftline{\twelvebf 5. Conclusions}
\vglue 0.4cm

We showed that holomorphy is the principle underlying the
non-renormalization theorem.  It enables us to control supersymmetric
strongly coupled field theories and to find exact results.

These strongly coupled theories exhibit a large number of interesting
phenomena:

\lfm{1.} Quantum moduli spaces

\lfm{2.} Smoothed out singularities

\lfm{3.} Non-conventional patterns of chiral symmetry breaking (not most
attractive channel)

\lfm{4.} Massless composite mesons and baryons

\lfm{5.} Lack of confinement and interacting CFT in four dimensions

\lfm{6.} Massless magnetic monopoles

\lfm{7.} Monopole condensation as mechanism for confinement

\lfm{8.} Calculable non-trivial terms in the effective Lagrangian

\lfm{9.} Calculable spectrum of stable particles

\medskip

We expect that further explorations of these theories will teach us even
more about the three applications we mentioned in the introduction
(dynamical supersymmetry breaking, topological field theories and the
dynamics of more generic strongly coupled four dimensional field theories).

\vglue 0.6cm
\leftline{\twelvebf Acknowledgements}
\vglue 0.4cm
I wish to thank the organizers of the conference for inviting me and for
arranging an interesting and stimulating meeting.  I also thank K.
Intriligator, R. Leigh and E. Witten for an enjoyable collaboration on
some of the work reported here.  Extremely useful conversations with T.
Banks, K. Intriligator, R. Leigh, G. Moore, R. Plesser, S. Shenker and
E. Witten are also acknowledged.  This work was supported in part by DOE
grant \#DE-FG05-90ER40559 and in part by NSF grant \#PHY92-45317.

\vglue 0.6cm
\leftline{\twelvebf References}
\vglue 0.4cm

\medskip

\itemitem{1.} N. Seiberg, \pl{318}{1993}{469}.
\itemitem{2.} N. Seiberg, \physrev{49}{1994}{6857}, hep-th/9402044.
\itemitem{3.} K. Intriligator, R. Leigh and N. Seiberg, hep-th/9403198,
RU-94-26, Phys. Rev. in press.
\itemitem{4.} N. Seiberg and E. Witten, hep-th/9407087,
RU-94-52, IAS-94-43; RU-94-60, IASSNS-HEP-94/55.
\itemitem{5.} V. Kaplunovsky and J. Louis, hep-th/9402005.
\itemitem{6.} I. Affleck, M. Dine, and N. Seiberg, \np{241}{1984}{493};
\np{256}{1985}{557}.
\itemitem{7.} E. Witten, \cmp{117}{1988}{353}; IASSNS-HEP-94-5,
hep-th/9403195.
\itemitem{8.} M.T. Grisaru, W. Siegel and M. Rocek,
\np{159}{1979}{429}.
\itemitem{9.} D. Amati, K. Konishi, Y. Meurice, G.C. Rossi and G.
Veneziano, \prep{162}{1988}{169} and references therein.
\itemitem{10.} M.A. Shifman and A.I Vainshtein, \np{277}{1986}{456};
\np{359}{1991}{571}.
\itemitem{11.} J. Polchinski and N. Seiberg, (1988) unpublished.
\itemitem{12.} E. Witten, \np{268}{1986}{79}.
\itemitem{13.} M. Dine and N. Seiberg, \prl{57}{1986}{2625}.
\itemitem{14.}A.C. Davis, M. Dine and N. Seiberg,
\pl{125}{1983}{487}.
\itemitem{15.}  V.A. Novikov, M.A. Shifman, A.I. Vainshtein and V.I.
Zakharov, \np{260}{1985}{157}.
\itemitem{16.} T. Banks, E. Rabinovici, \np{160}{1979}{349};
E. Fradkin and S. Shenker, \physrev{19}{1979}{3682}.
\itemitem{17.} N. Seiberg, to appear.
\itemitem{18.} S. J. Gates, Jr., \np{238}{1984}{349}; B. De Wit and
A. Van Proeyen, \np {245}{1984}{89};  S. Ferrara, \mpl{A6}{1991}{2175};
A. Strominger, \cmp {133}{1990}{163}; P. Candelas and X. de la Ossa,
\np {355}{1991}{455}.
\itemitem{19.} N. Seiberg, \pl{206}{1988}{75}.
\itemitem{20.} M. K. Prasad and C. M. Sommerfield,
\prl{35}{1975}{760}; E. B. Bogomol'nyi, {\twelveit Sov. J. Nucl. Phys.}
{\twelvebf 24} (1976) 449.
\itemitem{21.} E. Witten and D. Olive, \pl {78}{1978}{97}.
\itemitem{22.} E. Witten, \np{202}{1982}{253}.

\bye